\documentclass[runningheads]{llncs}
\usepackage{color}
\usepackage{amssymb}
\usepackage{soul}
\usepackage{comment}
\usepackage{multirow}
\usepackage{subcaption}
\usepackage{amsmath}
\usepackage{float}
\definecolor{mypink1}{rgb}{1, 0.5, 0.0}

\newcommand{\norm}[1]{\left\lVert#1\right\rVert}
\usepackage[utf8]{inputenc}
\usepackage{amsmath}
\usepackage{graphicx}
\usepackage{wrapfig}
\usepackage{caption} 
\emergencystretch=\maxdimen
\authorrunning{}% Part of LEFT running header
\titlerunning{}% Part of RIGHT running header

\begin{document}
\title{End-To-End Convolutional Neural Network for 3D Reconstruction of Knee Bones From Bi-Planar X-Ray Images}
\titlerunning{End-To-End CNN for 3D Reconstruction of Knee Bones From X-Ray Images}

\author{Yoni Kasten*\inst{1,2} \and
Daniel Doktofsky* \inst{1} \and
Ilya Kovler*\inst{1}}
\authorrunning{RSIP Vision}
\institute{RSIP Vision \and
Weizmann Institute of Science}

\maketitle              % typeset the header of the contribution
\setcounter{footnote}{0}
\begin{abstract} We present an end-to-end Convolutional Neural Network (CNN) approach for 3D reconstruction of knee bones directly from two bi-planar X-ray images. Clinically, capturing the 3D models of the bones is crucial for surgical planning, implant fitting, and postoperative evaluation. X-ray imaging significantly reduces the exposure of patients to ionizing radiation compared to Computer Tomography (CT) imaging, and is much more common and inexpensive compared to Magnetic Resonance Imaging (MRI) scanners. However, retrieving 3D models from such 2D scans is extremely challenging. In contrast to the common approach of statistically modeling the shape of each bone, our deep network learns the distribution of the bones' shapes directly from the training images. We train our model with both supervised and unsupervised losses using Digitally Reconstructed Radiograph (DRR) images generated from CT scans. To apply our model to X-Ray data, we use style transfer to transform between X-Ray and DRR modalities. As a result, at test time,  without further optimization, our solution directly outputs a 3D reconstruction from a pair of bi-planar X-ray images, while preserving geometric constraints. Our results indicate that our deep learning model is very efficient, generalizes well and produces high quality reconstructions.
\end{abstract}
\makeatletter
\def\blfootnote{\xdef\@thefnmark{}\@footnotetext}
\makeatother
\blfootnote{*Equal contributors}
\blfootnote{This work has been done at RSIP Vision}
\begin{figure}[tp]
    \centering
    \includegraphics[width=0.7\textwidth]{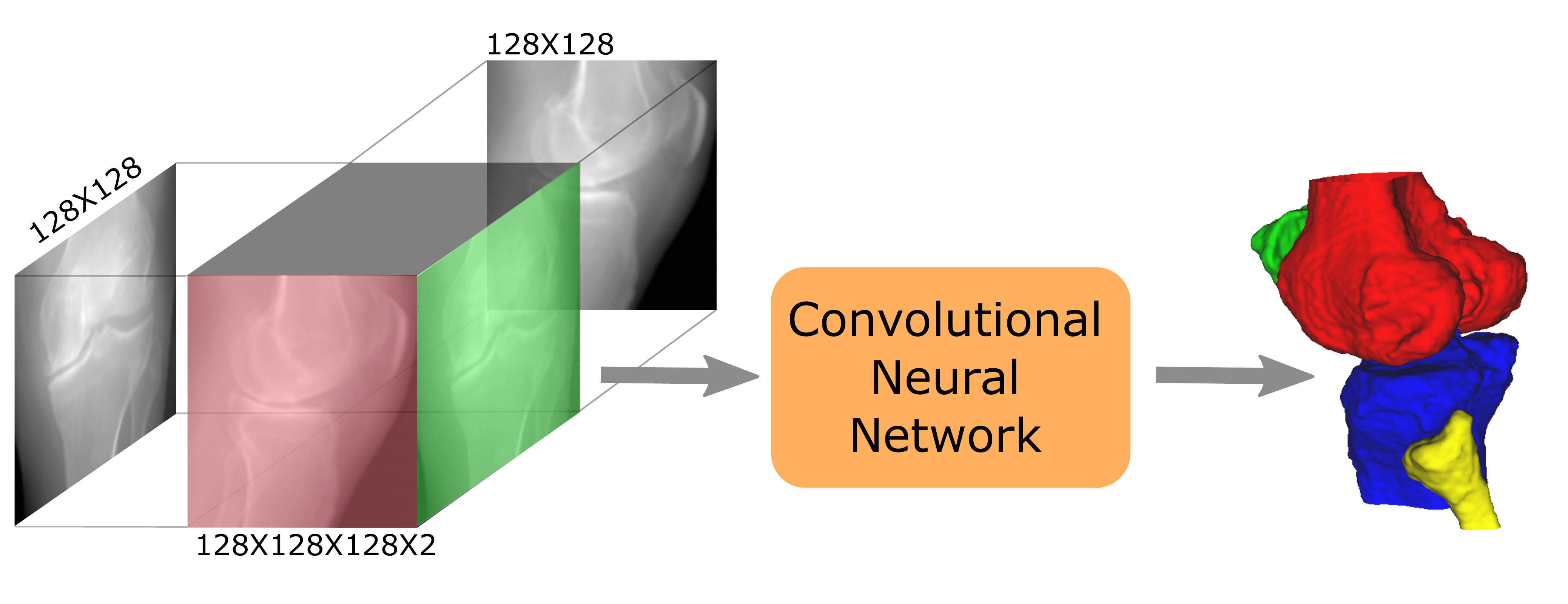}
    \caption{General scheme: AP and lateral X-ray scans of the knee joint are replicated into a 3D array, each on a different channel (green and red for illustration). A CNN then predicts a 3D segmentation map of the bone classes which is used for 3D reconstruction of the bones.}
    \label{fig:scheme}
\end{figure}
\vspace{-6mm}
\section{Introduction }
3D reconstruction of knee bones is an important step for various clinical
applications. It may be used for surgical planning, precise implant selection, patient specific
implant manufacturing or intraoperative jig printing which perfectly fits the
anatomy. X-ray images are often used due to their wide availability, lower
price, short scanning time and lower levels of ionizing radiation compared to CT scanners.
However, since X-ray images provide only 2D information, some prior
knowledge must be incorporated in order to extract the missing dimension.
Previous approaches
 \cite{cootes1995active,lamecker2006atlas,fotsin2019shape,ehlke2013fast} use Statistical Shape Models (SSM) or Statistical
Shape and Intensity Models (SSIM) for reconstructing bones from X-ray
images. However, optimization for the deformable model parameters might
be slow and needs a good initialization point to avoid local maxima 
 \cite{reyneke2018review,kim20193d}. 

In this paper, we present a novel end-to-end deep learning approach for 3D
reconstruction of knee joints from two bi-planar X-ray images. The overall
scheme is presented in Fig.~\ref{fig:scheme}. CNNs have recently proven 
very effective for various types of tasks 
 \cite{lecun2015deep}, including image segmentation and
classification. However, implementing 3D reconstruction from two or more 2D images using a deep learning approach remains a challenging task,
 due to the difficulty
of representing a dimensional enlargement in multi-view settings with standard
differentiable layers. Moreover, due to the transparent nature of X-ray images, matching surface points between
multi-views for dense reconstruction is extremely challenging compared to the standard
multi-view setting
\cite{hartley2003multiple}.

We address these challenges by introducing a dimensional enlargement
approach that given two bi-planar X-rays back-projects each pair of
corresponding epipolar lines into a two-channeled epipolar plane. This results
in a 3D volume that contains all the information observed from the two X-ray
images, while preserving the two-view geometric constraints. We combine this
representation with a deep learning architecture that outputs 3D models of the
different bones. The experiments show the utilization of our approach for 3D
reconstruction of knee joint. We strongly believe that our method paves the
way for future research in deep learning based 3D modeling of bones from
X-ray scans. In contrast to SSM based methods, our method does not require
an initialization and runs in $0.5$ seconds while a standard SSM
optimization for one knee bone takes about $4.88$ seconds.

%The structure of the rest of the paper is as follows: In Sec.\ref{sec::rel_work} we give an overview of related works, in sec.3 we describe the method \yk{and so on}

\section{Related work}
\label{sec::rel_work}

\noindent \textbf{Deformable models} 3D bones reconstruction from X-ray images is mostly done by SSM \cite{cootes1995active,lamecker2006atlas,cresson20103d,baka20112d} for bones surface modeling, or SSIM\cite{fotsin2019shape,ehlke2013fast} for further modeling bones interior density. We refer the readers to \cite{reyneke2018review,goswami20153d} for comprehensive overviews of the existing methods. The basic principle is to rigidly align a collection of 3D models and to characterize their non-rigid mutual principal components. Then, given one or more X-ray images, a 3D reconstruction of the bones is achieved by optimizing the model parameters to maximize the similarity between its rendered versions to the input X-ray images. Recently, \cite{kim20193d}  used a deep learning approach for detecting landmarks in X-ray images and triangulating them to 3D points. However, their network does not directly outputs bone reconstructions, and the detected 3D landmarks are only used to initialize a 3D deformable model. As a result, an SSM optimization is still required and takes around 1 minute.

\noindent \textbf{Reconstruction from multiple images} %While reconstruction of shape from a single image is a geometrically ill-posed problem, 
Several recent methods \cite{wang2018pixel2mesh,wu2017marrnet} use deep learning approaches for reconstructing a shape from single image, for predefined objects \cite{wu20153d}. Geometrically, by using two or more images, it is possible to reconstruct a 3D surface by triangulating corresponding points, assuming the cameras' relative positions are known \cite{hartley2003multiple}. The relative poses of the cameras can be computed by matching points \cite{lowe2004distinctive} or lines  \cite{Wurfl_2019_ICCV,kasten2016fundamental,ben2016camera} descriptors.  Several recent papers use deep learning approaches to reconstruct shapes from two and more images.  3D-R2N2 \cite{choy20163d} and LSM \cite{kar2017learning} use RNNs to fuse feature from multiple images for reconstructing a binary voxels mask for representing 3D models. In contrast,   \cite{xie2019pix2vox} reconstructs one volume from each image and fuses them in a context-aware layer.  \cite{wen2019pixel2mesh++} initializes a mesh from one of the views by \cite{wang2018pixel2mesh} and refines it by repeatedly applying graph convolutional layers on its 3D coordinates with learned 2D features sampled from the projections of the 3D points on the multiple images. 
\cite{chen2019using} uses deep network to reconstruct  3D models from simulated bi-planar X-ray images of a single spine vertebra by applying 2D convolutional layers to encode the images into a feature vector, and then decoding it to a 3D reconstruction using 3D convolutional layers. In contrast, our method uses more effective, and geometrically consistent network architecture that uses end-to-end 3D convolutional layers with skip connections, enabling faster and more accurate reconstruction of multi-class bones as we show in Sec. \ref{sec::real_expr}. 

\noindent \textbf{Computed Tomography from X-ray images} Although mathematically, generation of computed tomography from few images is an ill-posed problem, a prior knowledge on the scanned objects can approximate the free parameters.    
X2CTGAN \cite{ying2019x2ct} uses an end-to-end deep learning approach for reconstructing a CT from X-ray images. \cite{shen2019patient} trains a patient-specific deep network to extract a CT volume from single X-ray image.  \cite{henzler2018single} uses a deep network to reconstruct computed tomography of different mammalian species from single X-ray images. However, these approaches only estimate CT volumes, and another challenging segmentation step is required for extracting  3D reconstructions of the anatomical objects.

%\begin{figure}
 %   \centering
  %  \includegraphics[width=0.35\textwidth]%{labels.pdf}
%     \caption{Class labels of knee joint: 0 - background, 1 - Femur, 2 - Patella, 3 - Tibia, 4 - Fibula}
%     \label{fig:labels}
%\end{figure}

\section{Method}
\label{sec::method}
\begin{figure}[t]
    \centering
    \includegraphics[width=0.9\textwidth]{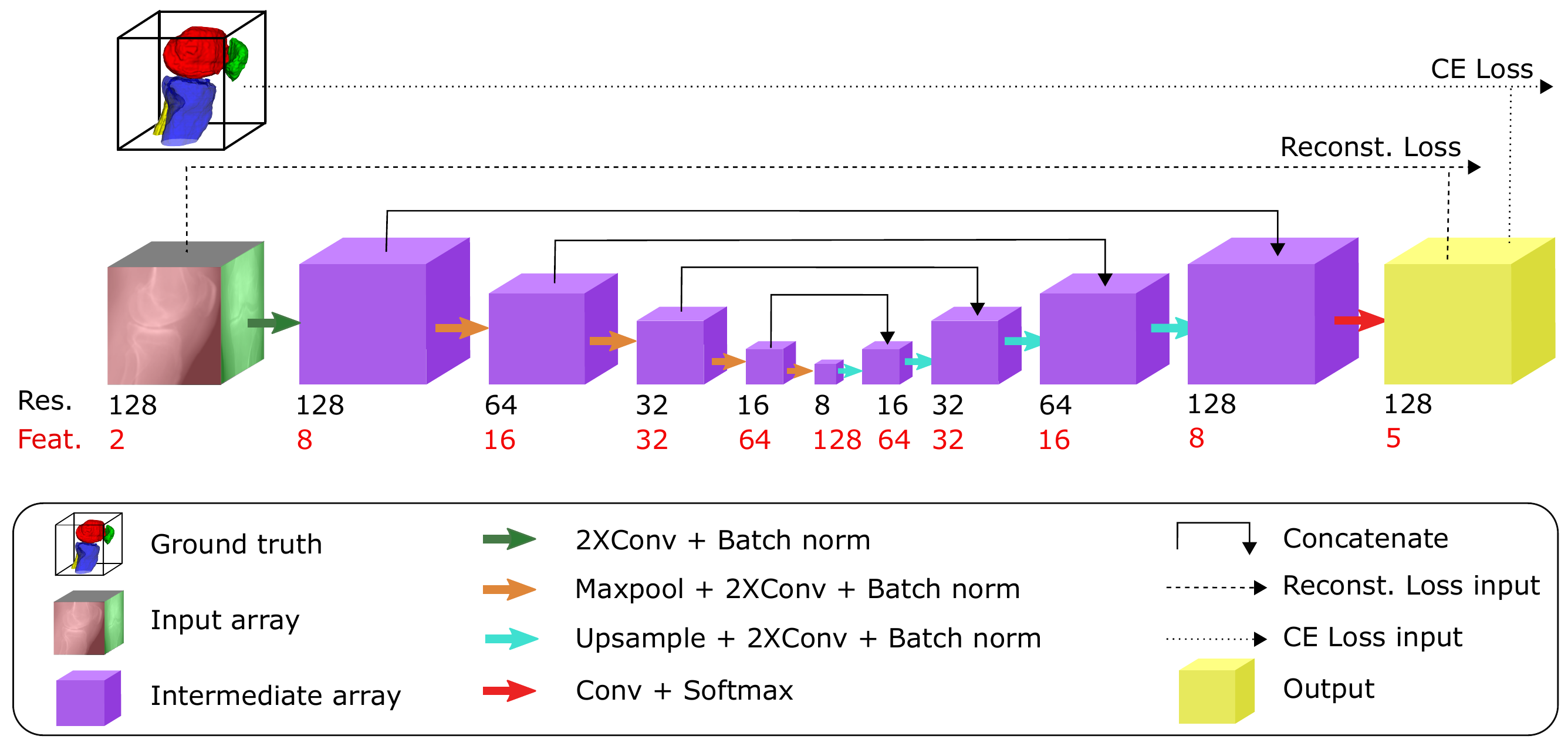}
    \caption{Our deep network architecture and loss functions as described in Sec.~\ref{sec::method}.}
    \label{fig:unet}
    \vspace{-2mm}
\end{figure}

\subsection{Network Architecture}\label{sec::architecture}
%The overall architecture of our deep network is presented in Fig.\ref{fig:unet}. The network has two main parts: dimensional enlargement, and a U-Net architecture of 4 levels.
%\yk{Old:}
%Given two bi-planar X-ray images of lateral and Anterior-Posterior (AP) views both of sizes $128 \times 128$, we assume orthogonal projection camera models. Furthermore, assuming that the two images are rectified, that is for $i= 1,\dots, 128$ the $i^{th}$ rows of the two images are two corresponding epipolar lines, we generate a two channeled volume by replicating the AP and lateral X-ray 2D-scans over the 0 and 1 axes, respectively. This volume representation is geometrically consistent with the input images as follows: each input X-ray image forms a DRR from this volume through the corresponding axis and channel.

%\yk{Suggested:}
Given two bi-planar X-ray images of lateral and Anterior-Posterior (AP) views, both of sizes $128 \times 128$, we first create a two channeled volume representation of size $128\times 128 \times 128$. As illustrated in Fig.\ref{fig:scheme}, the volume has two channels, each contains one view ( lateral or AP) replicated $128$ times over one dimension (0,1 respectively). Assuming that the input images are rectified orthographic projections from orthogonal views, each axial slice in this volume contains an epipolar plane, with voxels, back-projected from pixels of two corresponding epipolar lines. Therefore, this 3D representation is geometrically consistent with the input images.    

The rest of the architecture is inspired by \cite{milletari2016v} and presented in Fig.~\ref{fig:unet}. We use 3D convolutions of  size $3\times3\times 3$, and skip connections  between the encoding and the decoding layers. The last layer is a $1\times 1\times 1 $ convolution block with 5 output channels, representing 5 output classes followed by a Softmax activation. Classes $0$-$4$ represent an anatomical partitioning of the knee bones (see Fig.~\ref{fig:qual_synth}(e)).  

\vspace{-4mm}
\subsection{Training}\label{sec::training}
While CT images with ground truth 3D segmentation are available, pairs of X-ray images with associated ground truth 3D reconstructions are very rare. Moreover, geometrical alignment of  each ground truth reconstruction with its X-ray images requires a 2D-3D registration process, which is itself challenging and error prone.  Instead, we use annotated CT scans to create synthetic X-ray images by rendering DRRs. This way, each pair of synthetic X-ray images is associated with an aligned ground truth reconstruction.

\begin{figure}[t!]
    \centering
    \includegraphics[width=0.8\textwidth]{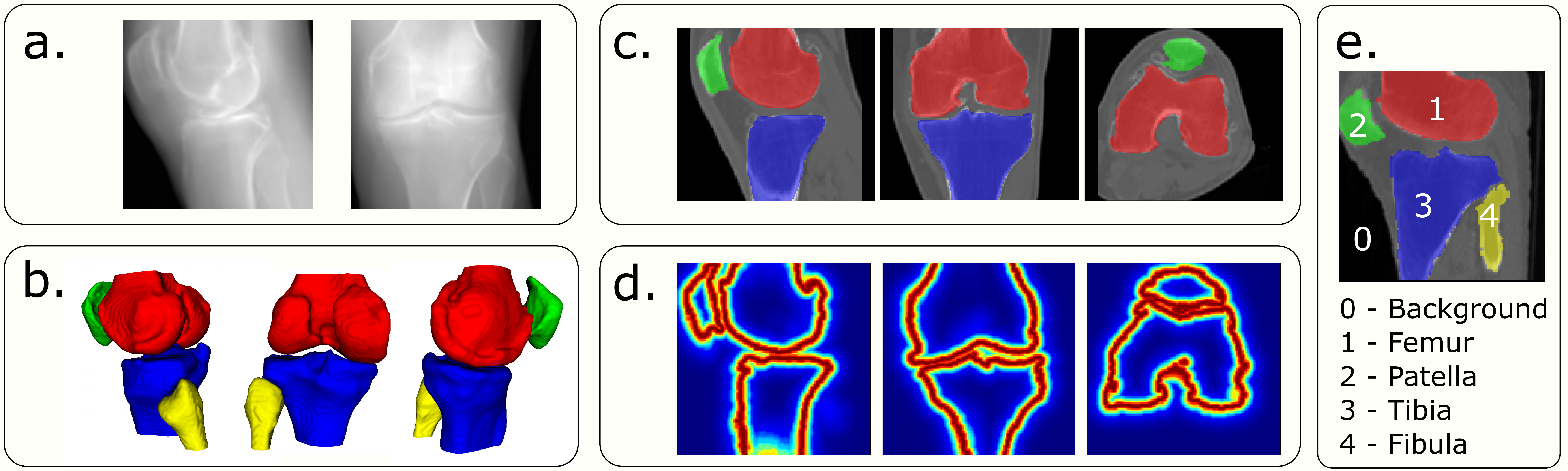}
    \caption{Test set qualitative results for DRR input images (a-c), and training data visualization(d-e). (a) Biplanar input DRRs from the Test set. (b) Our reconstruction result from different view points. (c) Our reconstruction result displayed over a referenced CT scan, on 3 different axes. (d)  Slices of Distance Weight Map (DWM) on 3 different axes. (e) Bone types and their assigned labels.}
    \label{fig:qual_synth}
    \vspace{-2mm}
\end{figure}

For a supervised loss function, inspired by  Fidel et al. \cite{guerrero2018multiclass} we spatially weight the cross-entropy loss to give more importance to the challenging near surface voxels. For each training sample, we define a spatial 3D Distance Weight Map (DWM) that has a size of the ground truth volume where its value on voxel $i$ is defined by:
\begin{equation}DWM(i) = 1 + \gamma\cdot exp(-d(i)/\sigma)\end{equation}
Where $d$ is a distance transform that  specifies for each voxel its corresponding distance from any bone surface, and  $\gamma,\sigma$ are constants which we set to $8,10$ respectively for all the training samples. A visual example is presented in Fig.~\ref{fig:qual_synth}(d). The DWM is then applied for weighting the voxel-wised cross entropy loss as follows: 
\begin{equation}loss_{CE} = -\frac{1}{N}\sum_{i=1}^N\sum_{k=0}^4 DWM(i)\cdot q_{k}(i)\cdot log(p_{k}(i))
\end{equation}
where $i$ is the index of a voxel, $N$ is the total number of voxels, $k$ is the class label, $q_k(i)\in \{0,1\}$ and $p_k(i)\in (0,1)$ are respectively the ground truth and network prediction probabilities of voxel $i$ being labeled $k$.

We further define an unsupervised reconstruction loss to align the network prediction of bones probability map with the input X-ray images. Even though the input X-ray images contain bones together with additional anatomical elements, the image gradients of the bones are quite dominant. Therefore, the input X-rays are expected to have gradients that are relatively correlated  with the DRRs from the predicted bones probability map.

The reconstruction loss is defined by:  \begin{equation}Loss_{reconst} = 1 -\frac{1}{2}\left(NGCC(I_{Lat},DRR_{Lat}) +(NGCC(I_{AP},DRR_{AP})\right) \end{equation}
where NGCC is the Normalized Gradient Cross Correlation\footnote{The exact definition is given in the supplementary material}, $I_{AP},I_{Lat}$ are the input X-ray images from AP and lateral views respectively, and $DRR_{AP},DRR_{Lat}$  are DRRs applied on the maximum over the bones channels of the network prediction. This loss encourages the network to use the available information of the input images, that can actually be used in inference time, where no supervision is available.  We observe that this loss improves the generalization of the network to unseen images (see 
Sec.~\ref{sec::synthetic_expr}). Overall our loss function is: \begin{equation}Loss = \frac{1}{2}(Loss_{reconst} + Loss_{CE})\end{equation} For training the network, Adam optimizer was used with initial learning rate of $10^{-2}$, divided by a factor of $10$  every $10$ epochs.   We used training,validation and test sets of $188$,$10$ and $20$ scans respectively, created from knee joint CT scans with associated GT segmentations and reconstructions. Each scan was augmented by rotating it randomly with random angles of range $(-5,5)$ and projected into 2 bi-planar DRRs which are used as synthetic input X-rays. We trained the network for 23 epochs. 

  \begin{table}[t]
\centering
\caption{Evaluation metrics for our results given inputs of bi-planar DRRs. The results are averaged over the test set of unseen $20$ scans.}
\begin{tabular}{|| c |c | c| c| c |c |c||} 
 \hline
  & Background & Femur & Patella & Tibia & Fibula &  Bones average\\ [0.5ex] 
 \hline\hline
% Hausdorff & - & {3.804}  & {3.166} &{4.467} &  3.288 & 3.68\\
    %  \hline
Chamfer(mm)& - &  {1.075 }  & {1.709} & {1.175} & {1.218}& 1.294\\
        \hline
  Dice & 0.986 &  {0.943} & {0.894} & {0.945} &{0.848} & 0.907\\ 
  \hline

 \end{tabular}
  \label{tab::drr_metrics}
  \vspace{-3mm}
\end{table} 
%\vspace{-3mm}
\subsection{Domain adaptation}\label{sec::adaptation}%\vspace{-1mm}
X-ray images have a different appearance than DRRs. In order to apply our deep model on X-ray images, we trained a network that is based on CycleGAN \cite{zhu2017unpaired} to transfer them to have a DRR-style appearance. During training, in each iteration the model uses two non aligned images $I_{Xray}$ and $I_{DRR}$ to generate two fake images: $I_{DRR\rightarrow Xray},I_{Xray\rightarrow DRR}$. In order to generate DRR-style images which are completely aligned with the input X-ray images we use the original CycleGAN with additional content preserving loss function: \begin{equation}L_{Cont} = 1 -\frac{1}{2}\left(ZNGCC(I_{Xray\rightarrow DRR},I_{Xray}) +(ZNGCC(I_{DRR\rightarrow Xray},I_{DRR})\right) \end{equation}
Where ZNGCC is the Zero Normalized Gradient Cross Correlation\footnote{The exact definition is given in the supplementary material}.
 We trained the style transfer model with training/validation sets of $370/57$ pairs of bi-planar X-ray images of the knee,  for $30$ epochs. In the supplementary material we show visual results of the style transfer process.
 
 \begin{table}[h!]
\centering
\caption{Quantitative evaluation of real $28$ test pairs of X-rays inputs, and comparisons with the femur reconstructions of \cite{klima2015gp} and \cite{chen2019using}. The results are averaged over the test set. The patella metrics computed only on the lateral view (GT annotations for the AP view are unavailable).}
 \begin{tabular}{||l |c| c|c|c| c| c | c |c||} 
 \hline
 &\multicolumn{2}{|c|}{Femur SSIM \cite{klima2015gp}} & \cite{chen2019using}& \multicolumn{5}{|c||}{Ours}  \\
 \cline{2-9}
  &\text{Manual}&\text{Perturbed}&Femur&Femur&Patella& Tibia& Fibula&Bones avg.\\ [0.5ex] 
 \hline\hline
 % Hausdorff &    39.892  & 65.582 &\textbf{17.67} & 15.527 &28.056 &  16.303 &19.38 & 84.86\\
  %    \hline
  Chamfer(mm)&  7.529  & 8.559& 3.984 & \textbf{1.691} & 1.198 & 1.135 & 2.873& 1.778 \\
        \hline
  Dice &    0.803 & 0.783&0.878 &\textbf{0.948} &  0.91 & 0.959 &0.809&0.906\\ 
  [1ex] 
 \hline
 \end{tabular}
  \label{table:result_table_xray}\vspace{-4mm}
\end{table}
 \begin{table}[t]
\centering
 \caption{Ablation study. Measuring the importance of different components of our model for real X-rays. The results metrics are averaged over the 4 reconstructed bones.}
\begin{tabular}{||l |c| c|c| c|c||} 
 \hline
  & Full &Without $DWM$  & Lateral only & No $Loss_{reconst}$  &No style transfer \\ [0.5ex] 
 \hline\hline
% Hausdorff & \textbf{3.68} & 6.01  & 4.66 & 3.83  \\
   %   \hline
Chamfer(mm)& \textbf{1.778} &  {1.863 }  & {2.979} & {1.92}&6.146 \\
        \hline
  Dice & \textbf{0.906} &  0.901 & 0.844 & 0.892&0.742 \\ 
  \hline

 \end{tabular}
  \label{tab::ablation}
  \vspace{-1mm}
\end{table} 
\section{Experiments}
\subsection{DRR inputs}\label{sec::synthetic_expr}
We tested our method on a test set of 20 scans (see  Sec.~\ref{sec::training}), and evaluated the results using the ground truth 3D segmentations and reconstructions. Each pair of bi-planar DRRs is used as an input to our deep network described in Sec.~\ref{sec::architecture}. For each testing sample we used the Marching Cubes algorithm\cite{lorensen1987marching} to extract a set of 3D bones meshes from the predicted volumetric labels. A qualitative result is presented in Fig.~\ref{fig:qual_synth}a-\ref{fig:qual_synth}c. Quantitative metrics are calculated for each bone type and presented in Table \ref{tab::drr_metrics}. Dice (higher is better) is computed over the predicted voxels maps and Chamfer (lower is better) is computed directly on the final reconstructions.

\vspace{-4mm}
\subsection{Real X-ray test cases}
\label{sec::real_expr}
\begin{figure}[t]
    \centering
      \includegraphics[width=\textwidth]{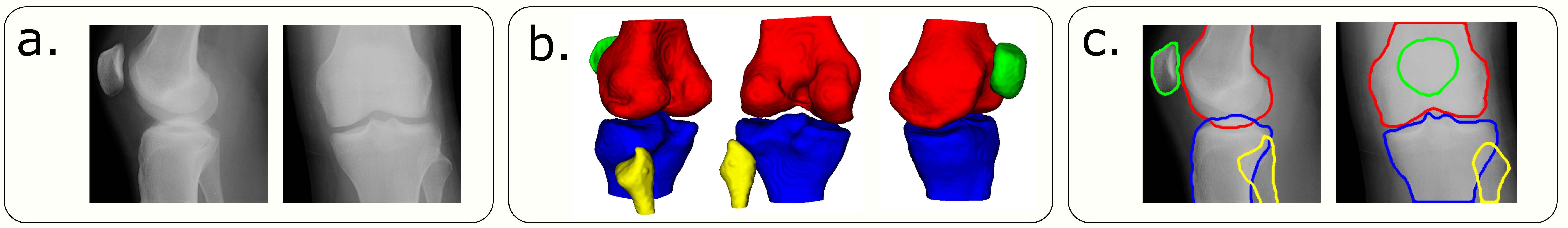}

    \caption{Qualitative results on real X-rays.  (a) Biplanar input X-rays. (b) 3D reconstruction result displayed from different view angles. (c) Boundaries of reconstruction projections displayed over the input X-rays .}
    \label{fig:xray_results}\vspace{-2mm}
\end{figure}
We evaluated $28$ test cases of X-ray images. Each pair of lateral and AP X-ray images was cropped manually by an expert to contain a bi-planar pair of rectified images of the knee joint such that, when resized to $128\times 128$ pixels, the pixel size is $1$ mm. Even though the view directions of the X-ray images are not guaranteed to be exactly orthogonal, our method, trained on purely orthogonal inputs, handled such cases successfully. We applied domain adaptation procedure to transform their style as described in Sec.~\ref{sec::adaptation}, and applied the network on the transformed X-ray images. % We further calibrated the X-rays as described on Sec.~\ref{sec::calibration}. 
In Fig.~\ref{fig:xray_results} we present a qualitative results of a 3D reconstruction given an input of bi-planar real X-ray images.  Since 3D ground truth is not available for the X-ray images, we use 2D bi-planar ground truth multi-class masks annotated by experts for each case for evaluation: each reconstructed 3D model is projected to the 2 X-ray views and the evaluation metrics are computed relative to the GT masks. 

We compare our performance with two baseline methods: the femur SSIM model\footnote{Only their femur model has an available code} of \cite{klima2015gp}, and the single bone reconstruction deep network of \cite{chen2019using}, trained using our training set for reconstructing the femur, and tested with the real X-ray test images (after applying our domain adaptation). Quantitative comparisons are presented in Table~\ref{table:result_table_xray}. Since \cite{klima2015gp} requires initialization for the SSIM model, we initialized it manually to the best of our ability. The optimization of \cite{klima2015gp} converged after $4.88$ seconds, while our method, without any initialization reconstructs $4$ bone types in $0.5$ seconds, and achieves better results. Our method is more accurate and much faster than \cite{chen2019using} which runs in $45$ seconds for one bone reconstruction. To demonstrate the initialization sensitivity of \cite{klima2015gp}, for each case of the test cases we applied a random perturbation on the manual initialization: we shifted the position parameters in a range of $20$mm, multiplied the scale parameters  by a factor of range $[0.985,1.015]$, and evaluated the average results (see Table~\ref{table:result_table_xray}, "Perturbed"). The average running time for the perturbed initialization increased from $4.88$ seconds to $6.05$ seconds, while 34\% of the perturbed cases did not converge at all.  We further show an ablation study in Table~\ref{tab::ablation} of running the model without several of its components to evaluate their importance. %We also performed an ablation study for running our method directly on the X-rays without style transfer. The Chamfer and Dice metrics averaged over the bones degraded significantly from 2.75mm and 0.84 to 32.8mm and 0.18 respectively.
\paragraph{Technical details}
We performed all of our experiments on a computer with MS Windows 10 64bit OS, Intel i7 7700K CPU and Nvidia GeForce GTX 1070 graphic card.
The data is provided by a third party who has obtained consent for use in research. 
\vspace{-2mm}
\section{Conclusion}
\vspace{-2mm}
We presented an effective end-to-end deep network for  knee bones 3D reconstruction from bi-planar X-ray scans. We used a novel representation, training from synthetic data and domain adaptation to achieve an efficient, robust and accurate method. In the future we would like to extend our method to more bones reconstruction setups, and to extend the geometric 2D-3D representation of our model for additional X-ray projection models.  

\noindent \footnotesize \textbf{Acknowledgements.} The authors thank Aliza Tarakanov, Arie Rond, Noy Moskovich, Eyal Shenkman, Eitan Yeshayahu, Yara Hussein, Astar Maloul-Zamir, Liam Simani, Polina Malahov and Amit Hadari M.D, for their help in datasets creation and annotation.
\clearpage
\bibliographystyle{splncs04}
\bibliography{bibfile}
\clearpage
\appendix
\section{Supplementary Material}
\begin{equation}\label{eq:gcc}NGCC(I_1,I_2)=\frac{1}{2}\left( \frac{{G_x(I_1)}}{\norm{{{G_x(I_1)}}}}\cdot \frac{{G_x(I_2)}}{\norm{{G_x(I_2)}}}+\frac{{G_y(I_1)}}{\norm{{{G_y(I_1)}}}}\cdot \frac{{G_y(I_2)}}{\norm{{G_y(I_2)}}}\right)\end{equation}
\begin{equation}
\overline{G_x(I)}=G_x(I)-\text{mean}(G_x(I)),\overline{G_y(I)}=G_y(I)-\text{mean}(G_y(I))\end{equation}

\begin{equation}\label{eq:gcc}ZNGCC(I_1,I_2)=\frac{1}{2}\left( \frac{\overline{G_x(I_1)}}{\norm{\overline{{G_x(I_1)}}}}\cdot \frac{\overline{G_x(I_2)}}{\norm{\overline{G_x(I_2)}}}+\frac{\overline{G_y(I_1)}}{\norm{\overline{{G_y(I_1)}}}}\cdot \frac{\overline{G_y(I_2)}}{\norm{\overline{G_y(I_2)}}}\right)\end{equation}

\begin{figure}[H]
    \centering
    \begin{subfigure}[b]{0.4\textwidth}
        \includegraphics[width=\textwidth]{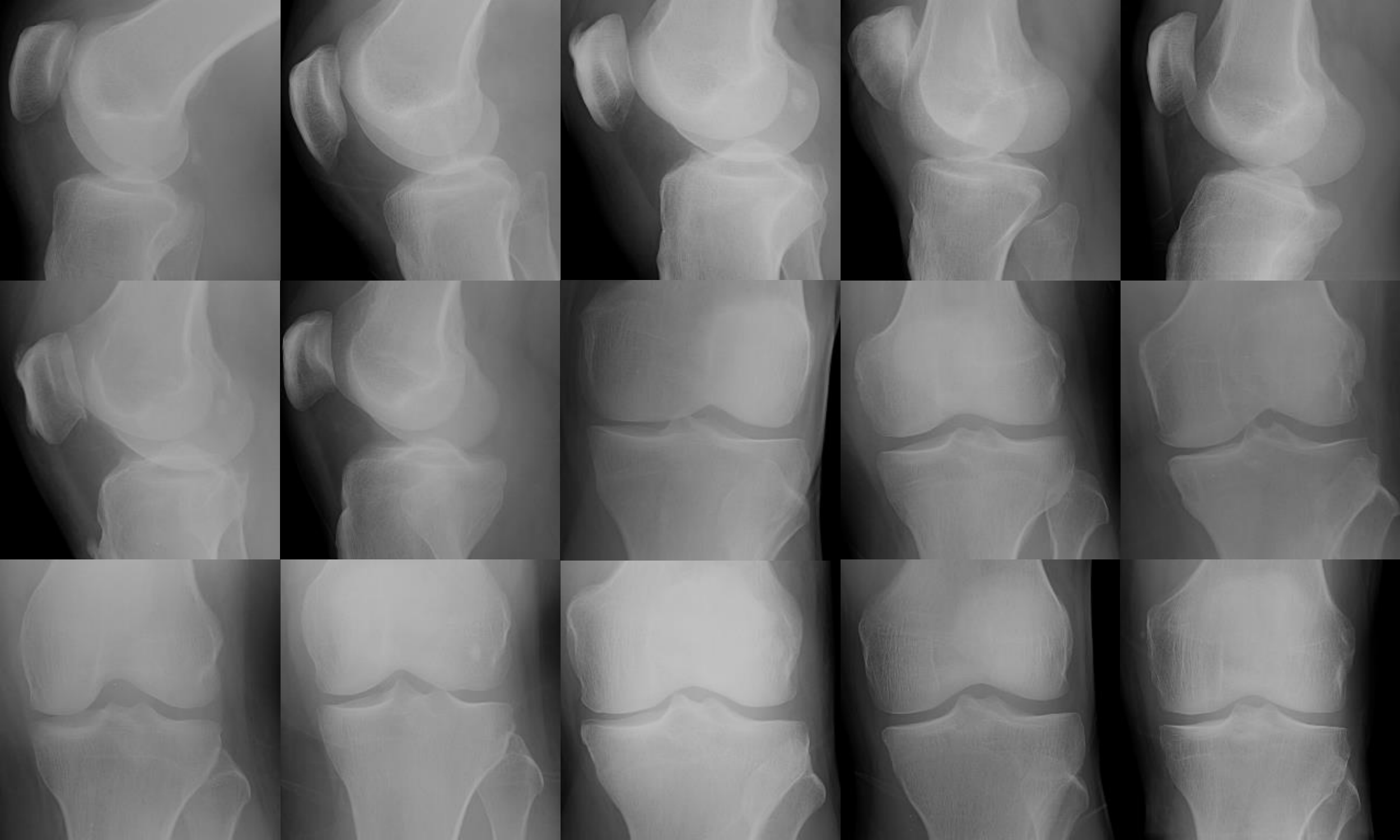}
        \caption{X-rays}
        \label{fig:gull}
    \end{subfigure}
    ~ %add desired spacing between images, e. g. ~, \quad, \qquad, \hfill etc. 
    \begin{subfigure}[b]{0.4\textwidth}
        \includegraphics[width=\textwidth]{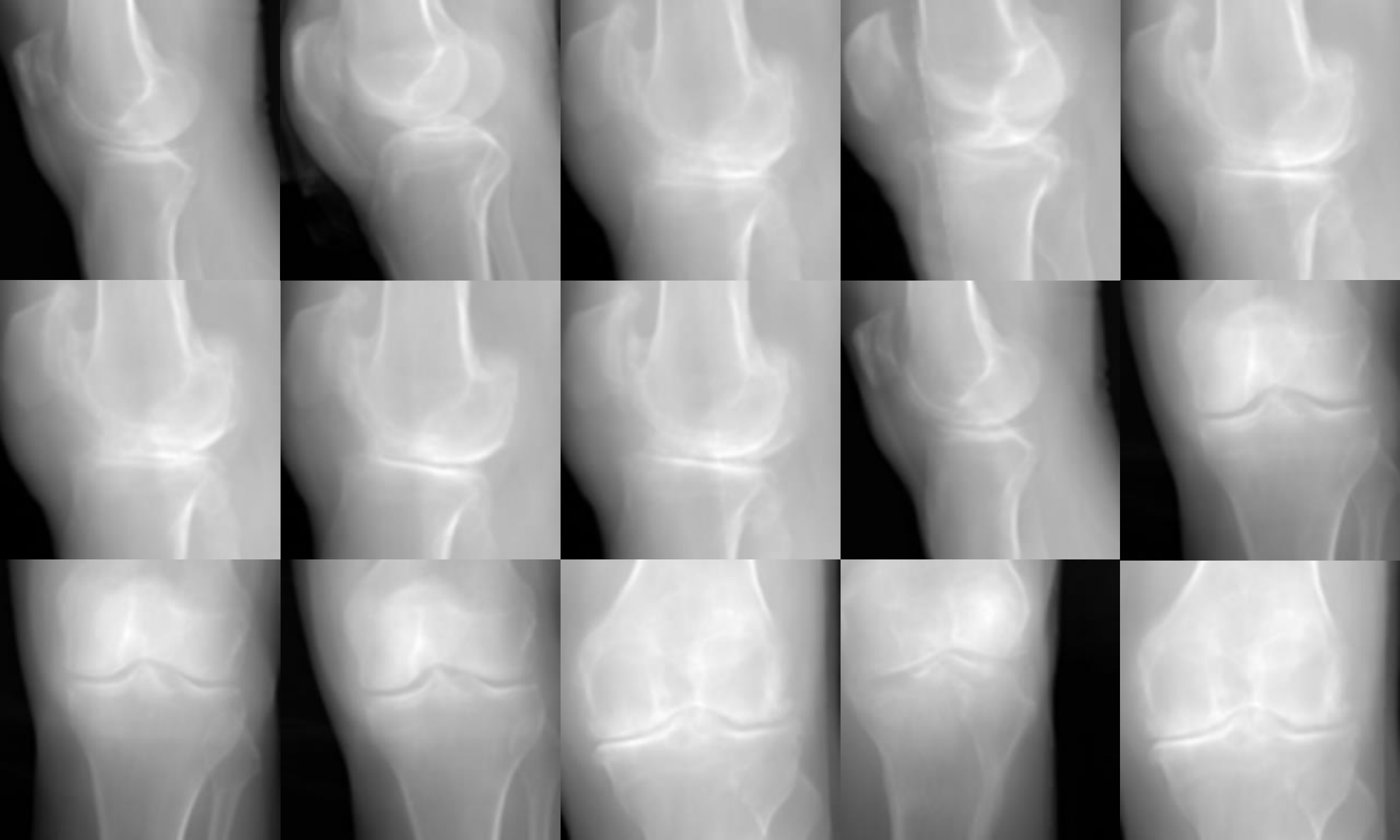}
        \caption{DRRs}
        \label{fig:tiger}
    \end{subfigure}
    ~ %add desired spacing between images, e. g. ~, \quad, \qquad, \hfill etc. 
  
    \caption{Visualization of the training data that we used for training CycleGAN. }\label{fig:animals}
\end{figure}

\begin{figure}[H]
    \centering
    \begin{subfigure}[b]{0.15\textwidth}
        \includegraphics[width=\textwidth]{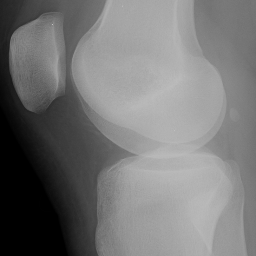}
        \includegraphics[width=\textwidth]{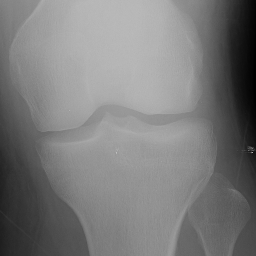}
        \caption{}
        \label{fig:gull}
    \end{subfigure}
    \ \ 
    ~ %add desired spacing between images, e. g. ~, \quad, \qquad, \hfill etc. 
    \begin{subfigure}[b]{0.15\textwidth}
        \includegraphics[width=\textwidth]{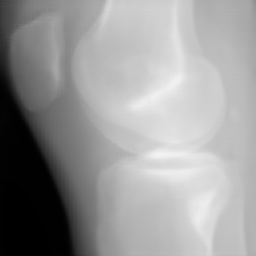}
        \includegraphics[width=\textwidth]{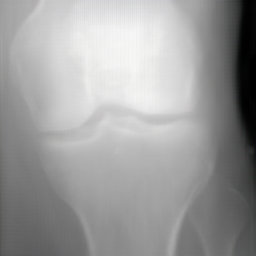}
        \caption{}
        \label{fig:tiger}
    \end{subfigure}
    \ \ 
    ~ %add desired spacing between images, e. g. ~, \quad, \qquad, \hfill etc. 
    \begin{subfigure}[b]{0.15\textwidth}
        \includegraphics[width=\textwidth]{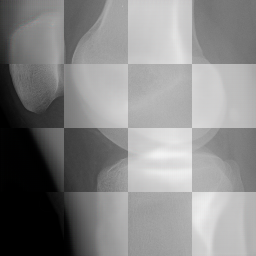}
        \includegraphics[width=\textwidth]{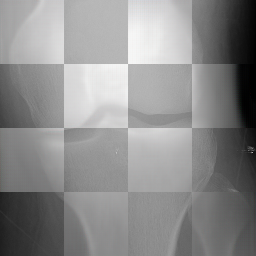}
        \caption{}
        \label{fig:mouse}
    \end{subfigure}
    \caption{CycleGAN results: (a) X-ray inputs. (b) DRR-style outputs. (c) Inputs-outputs content comparisons.}\label{fig:animals}
\end{figure}

\begin{figure}[t!]
    \centering
      \includegraphics[width=0.7\textwidth]{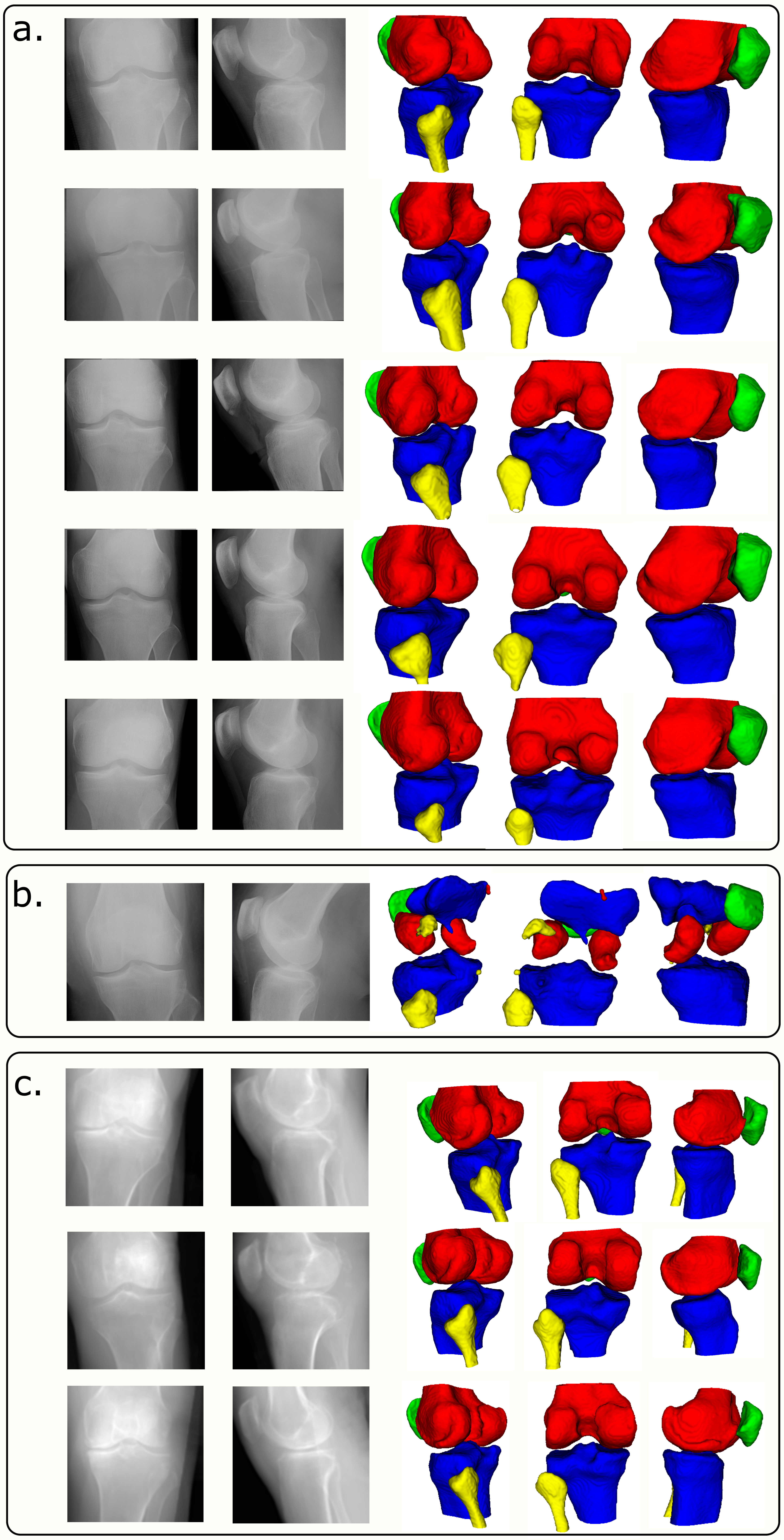}

    \caption{Supplementary qualitative results on test cases: left - inputs, right - outputs.  (a) Real X-rays inputs. (b) Real X-rays inputs without Cycle-GAN, as ablation study. (c)  DRRs inputs.}
    \label{fig:supp_quality_results}
\end{figure}

\end{document}